\documentclass[twocolumn,english,reprint, longbibliography, superscriptaddress, breaklinks=true, showkeys, showpacs=false, nofootinbib]{revtex4-1}
\usepackage[T1]{fontenc}
\usepackage[latin9]{inputenc}
\usepackage{color}
\usepackage{babel}
\usepackage{amsmath}
\usepackage{amssymb}
\usepackage{subfigure}
\usepackage{graphicx}
\usepackage{physics}
\usepackage{tikz}
\usepackage{hyperref}
\hypersetup{
    colorlinks=true,
    linkcolor=blue,
    filecolor=blue,      
    urlcolor=blue,
    }
\makeatletter
\@ifundefined{textcolor}{}
{%
 \definecolor{BLACK}{gray}{0}
 \definecolor{WHITE}{gray}{1}
 \definecolor{RED}{rgb}{1,0,0}
 \definecolor{GREEN}{rgb}{0,1,0}
 \definecolor{BLUE}{rgb}{0,0,1}
 \definecolor{CYAN}{cmyk}{1,0,0,0}
 \definecolor{MAGENTA}{cmyk}{0,1,0,0}
 \definecolor{YELLOW}{cmyk}{0,0,1,0}
}
\pdfoutput=1
\hypersetup{colorlinks=true,citecolor=blue,linkcolor=cyan,urlcolor=blue,filecolor= green, breaklinks=true}
\usepackage{url}
\usepackage{breakurl}
\makeatother
\usepackage{qcircuit}

\begin{document}
%%%%%\--------------------------------------------------------\%%%%%
%%%%%\----------\Elementos Pre-Textuais\----------------------\%%%%%
%%%%%\--------------------------------------------------------\%%%%%
\title{Influence of spin on tunneling times in the super-relativistic regime}

\author{Said Lantigua}
\email[]{said.lantigua@acad.ufsm.br}
\affiliation{Physics Departament, Center for Natural and Exact Sciences, Federal University of Santa Maria, Roraima Avenue 1000, 97105-900, Santa Maria, RS, Brazil}

\author{Jonas Maziero}
\email[]{jonas.maziero@ufsm.br}
\affiliation{Physics Departament, Center for Natural and Exact Sciences, Federal University of Santa Maria, Roraima Avenue 1000, 97105-900, Santa Maria, RS, Brazil}

%%%%%\--------------------------------------------------------\%%%%%
%%%%%\----------\Resumo\--------------------------------------\%%%%%
%%%%%\--------------------------------------------------------\%%%%%
\begin{abstract}
For the relativistic tunneling effect described using Dirac's equation, in [Phys. Rev. A 70, 052112 (2004)] the authors presented the deduction of a general result that allows for the determination of the phase time (group delay) as the sum of the particle dwell time inside a potential barrier and of the self-interference delay associated with the incident and reflected wave functions interaction. 
%However, although their results obtained for the relativistic regime are correct, there are some errors in the results presented when the non-relativistic limit is considered. 
In this article, 
%besides correcting these errors, 
a mathematical model is derived through a construction analogous to the proposal mentioned above, but based on an alternative representation for Dirac's equation. This representation is similar to the one introduced in [Found. Phys. 45, 1586 (2015)]. Thus, from the application of this model in the study of the tunneling effect in the absence of an external magnetic field, the influence of spin on the tunneling times is described. More specifically, the tunneling time is obtained as the sum of the dwell times inside the potential barrier for particles with spin up and spin down and the self-interaction time associated with the incident and reflected wave functions for particles with spin up.
\end{abstract}


%For the relativistic tunneling effect described using Dirac's equation, in [Phys. Rev. A 70, 052112 (2004)] the authors presented the deduction of a general result that allows for the determination of the phase time (group delay) as the sum of the particle dwell time inside a potential barrier and of the self-interference delay associated with the incident and reflected wave functions interaction. However, although their results obtained for the relativistic regime are correct, there are some errors in the results presented when the non-relativistic limit is considered. In this article, besides correcting these errors, a mathematical model is derived through a construction analogous to the proposal mentioned above, but based on an alternative representation for Dirac's equation. This representation is similar to the one introduced in [Found. Phys. 45, 1586 (2015)]. Thus, from the application of this model in the study of the tunneling effect in the absence of an external magnetic field, the influence of spin on the tunneling times is described. More specifically, the tunneling time is obtained as the sum of the dwell times inside the potential barrier for particles with spin up and spin down and the interaction time associated with the incident and reflected wave functions for particles with spin up.% We thus describe the influence of spin on tunneling times, even in the absence of an external magnetic field.
%%%%%\--------------------------------------------------------\%%%%%
\maketitle
%%%%%\--------------------------------------------------------\%%%%%
%%%%%\----------\Sec.I:Introdução\----------------------------\%%%%%
%%%%%\--------------------------------------------------------\%%%%%
\section{Introduction}
The unusual behavior of saturation in the tunneling times of non-relativistic particles in scattering problems, uncovered by Hartman in 1962 \cite{Hartman196200}, is known in the literature as the Hartman effect (HE). The discovery of the HE stimulated an intense debate about what is the correct definition for the travelling time when such incident particles manage to cross the potential barrier, and many related investigations were carried out \cite{Bandopadhyay200400,Bandopadhyay202101,Chuprikov200900,Goldberg196700,Hasan202100,Hasan202000,Longhi200100,Longhi202200,Martinez200600,Sokolovski200000,Winful200200,Winful200300,rivlin,rivlin2,landauer,petersen,muga,leavens,landauer2,hagmann}. The definitions obtained from the direct application of the Schr\"{o}dinger equation in these situations would lead to the apparent paradox of particles tunneling at hyper-luminous speeds \cite{Leavens198900,Delgado200300}. 
%Because of this unexpected prediction, this effect has received much attention in the literature 
%However, it is worthwhile mentioning that there is also an alternative interpretation that attributes such a saturation effect to the stored energy or to the number of particles in the barrier \cite{Winful200300,Winful200200}.
    
In fact, the two most acceptable tunneling times definitions in the literature are the dwell time, that is defined as the integral of the probability density within the potential barrier, and the phase time, that is defined as the variation of the transmitted phase with respect to the energy \cite{Hauge198700,Winful200301}. 
By virtue of such definitions, Winful, Ngom, and Litchinitser generalized the previous tunneling time definitions by deducing an exact relation that defines the phase time, for relativistic particles that satisfy Dirac's equation (DE), as the sum of the dwell time and of the self-interaction time between the incident and reflected wave packets \cite{Winful200400}. In the non-relativistic limit, their expressions lead to the results obtained from the direct application of the Schr\"{o}dinger equation \cite{Winful200301}. In these works, it is indicated the existence of saturation of the tunneling times with the width of the barrier, 
extending Hartman's prediction also to the relativistic regime.
% which implies that these times are not appropriate to correctly define the transit time scales in the propagation of the particle through the potential barrier.
    
%Consequently, 
But accepting such definitions as transit time scales would imply in accepting that the particle would somehow be aware of the increase in barrier width. The particle would increase its speed by just the right amount to be able to cover a greater distance in the same time. A statement that in the context of quantum mechanics seems untenable.     
As a result, Winful and coworkers made the suggestion to consider the width of the wave packet as a significant quantity to characterize the transit time in quantum tunneling. The wave packet can be much larger than the width of the barrier, which translates into a loss of sense of the notion of the transit time of any individual particle through the potential barrier when the uncertainty of its position is greater than the width of the barrier itself, a situation that can arise in quantum tunneling \cite{Winful200400,Winful200600}.
    
An alternative view for getting a correct definition of such timescales was developed in the works of Buttiker \cite{Buttiker198300} and R\'amos et al. \cite{Ramos202000}, where definitions are presented that allow for the determination of the tunneling time as a function of the electron spin evolution when crossing a potential barrier in the presence of a magnetic field and the experimental measurement of such a tunneling time, respectively. 
However, once one can not apply this technique for arbitrary systems, it cannot be taken as a general method for quantifying and measuring tunneling times.
Furthermore, in Ajaib's works \cite{Ajaib201500,Ajaib201600}, it was presented how the influence of spin can be obtained in scattering over potential barriers even without the presence of a magnetic field. In this article, this result is used to develop a model that allows us to extend the results obtained by Winful et al. and with that to determine the influence of the spin in the definitions of the tunneling times deduced from the sensitivity theorem of the wave function with respect to energy variations, but in the super-relativistic regime.
    
The remainder of this article is organized as follows. In Sec. \ref{sec:representation}, the deduction of an alternative representation of the DE is presented. Subsequently, in Sec. \ref{sec:sensitivity}, the deduction of the sensitivity theorem of the wave function to energy variations in the alternative representation is discussed. In Sec. \ref{sec:solution}, we introduce the solution of the DE and apply the expressions deduced in the previous section to determine the tunneling times in the lowest energy solution for relativistic and non-relativistic regimes in this alternative representation. In Sec. \ref{sec:spin}, by means of a procedure similar to the one applied in the previous section, we report the solution of the DE and the expressions for the tunneling times involving the influence of the spin, introduced in the alternative representation, in the super-relativistic regime and in the absence of an external magnetic field. Finally, in Sec. \ref{sec:conclusion}, we give our concluding remarks.
%%%%%\--------------------------------------------------------\%%%%%

%%%%%\--------------------------------------------------------\%%%%%
%%%%%\----------\Sec.II:DiracEAR\-----------------------------\%%%%%
%%%%%\--------------------------------------------------------\%%%%%
\section{Alternative Representation for Dirac's Equation}
\label{sec:representation}

In the works presented by Ajaib in 2015 and 2016 \cite{Ajaib201500,Ajaib201600}, an alternative representation for the Dirac equation was deduced in which the Schr\"{o}dinger-Pauli equation is contained when considering local invariance. Such equation allows for the description of how the spin of the particle affects the reflection and transmission coefficients when applied to quantum scattering problems. This stimulated us to apply this approach to study how the spin of the particles influences the tunneling time when they cross a constant potential barrier.
    
In this sense, the objective of this section is to determine a suitable alternative representation for the Dirac equation (DE) \cite{sakurai}:
\begin{equation}\label{e:ARDE1}
        \left\{ i \hbar \gamma^{\mu} \partial_{\mu} - i \gamma_{5} mc \right\} \psi (x^{\kappa}) = 0, \;\; \mbox{for} \;\; \mu,\kappa=0,1,2,3.
\end{equation}
Above $i=\sqrt{-1}$, $\hbar$ is Planck's constant, $m$ is the rest mass of the particle and $\gamma^{\mu}$ are the Dirac matrices
\begin{align}
    & \gamma^{0} = \alpha^{4} =
        \begin{pmatrix}
            \mathcal{I}_{2} & \mathcal{O}_{2} \\
            \mathcal{O}_{2} & - \mathcal{I}_{2}
        \end{pmatrix}, \label{e:ARDE2} \\
    & \gamma^{j} = \alpha^{4} \alpha^{j} =
        \begin{pmatrix}
            \mathcal{O}_{2} & \sigma^{j} \\
            -\sigma^{j} & \mathcal{O}_{2}
        \end{pmatrix}, \;\; \mbox{for} \;\; j=1,2,3, \label{e:ARDE3} \\
    & \gamma^{5} = i \gamma^{0} \gamma^{1} \gamma^{2} \gamma^{3} = - \frac{i}{4!} \varepsilon_{\mu \nu \kappa \lambda} \gamma^{\mu} \gamma^{\nu} \gamma^{\kappa} \gamma^{\lambda}, \label{e:ARDE4}
\end{align}
that satisfy relationships
\begin{equation}\label{e:ARDE5}
        \begin{aligned}
            \gamma^{\mu} \gamma^{\nu} + \gamma^{\nu} \gamma^{\mu} = 2g^{\mu \nu} \mathcal{I}_{4}, \;\; \mbox{for} \;\; \mu,\nu =0,1,2,3,
        \end{aligned}
    \end{equation}
and
\begin{equation}\label{e:ARDE6}
    \begin{aligned}
        (\gamma^{5})^{2} = \mathcal{I}_{4}, \;\; \gamma^{5} \gamma^{\mu} + \gamma^{\mu} \gamma^{5} = \mathcal{O}_{4}, \;\; \mbox{for} \;\; \mu =0,1,2,3.
    \end{aligned}
\end{equation}
Above 
\begin{equation}\label{e:ARDE7}
    \varepsilon_{\mu \nu \kappa \lambda} = 
    \left\{
        \begin{aligned}
            +1 & \;\; \mbox{if $(\mu, \nu, \kappa, \lambda)$ is an even permu-} \\
            & \;\; \mbox{tation of $(0,1,2,3)$}, \\
            -1 & \;\; \mbox{if $(\mu, \nu, \kappa, \lambda)$ is an odd permu-} \;\; & \\
            & \;\; \mbox{tation of $(0,1,2,3)$}, \\
            0 & \;\; \mbox{otherwise},
        \end{aligned}
    \right.
\end{equation}
is the Levi-Civita symbol in four dimensions, $g^{\mu \nu} = diag(1,-1,-1,-1)$ is the Minkowski metric, and $\mathcal{I}_{d}$ and $\mathcal{O}_ {d}$ is the $d\times d$ identity and null matrix, respectively, and $\sigma^{j}$ are the Pauli matrices:
\begin{equation}\label{e:ARDE8}
    \begin{split}
        \sigma^{1} =
        \begin{pmatrix}
            0 & 1 \\
            1 & 0
        \end{pmatrix}, \;\;
        \sigma^{2} &=
        \begin{pmatrix}
            0 & -i \\
            i & 0
        \end{pmatrix}, \;\;
        \sigma^{3} =
        \begin{pmatrix}
            1 & 0 \\
            0 & -1
        \end{pmatrix}.
    \end{split}
\end{equation}

The DE of Eq. (\ref{e:ARDE1}) can be rewritten as
\begin{equation}\label{e:ARDE9}
        \left\{ E \gamma_{0} - i \hbar c \gamma_{j} \partial_{j} - i \gamma_{5} mc^{2} \right\} \varphi (x^{l})
            = 0,
\end{equation}
with $j,l=1,2,3$. Following a procedure similar to that applied by Ajaib \cite{Ajaib201500,Ajaib201600}, we can obtain an alternative representation for the one-dimensional DE,
\begin{equation}\label{e:ARDE10}
    \gamma_{0} = \beta = \frac{ \eta + \eta^{\dagger} }{\sqrt{2}}, \;\; \gamma_{3} = \pm \alpha_{3} = \eta^{\dagger} \eta - \mathcal{I}_{4}
\end{equation}
and
\begin{equation}\label{e:ARDE11}
    i\gamma_{5} = \frac{ \eta - \eta^{\dagger} }{\sqrt{2}},
\end{equation}
using the matrix
\begin{equation}\label{e:ARDE12}
    \begin{split}
        \eta &= \frac{1}{\sqrt{2}}
        \begin{pmatrix}
            \mathcal{I}_{2} &   \sigma_{3}    \\
            -\sigma_{3}    & - \mathcal{I}_{2}
        \end{pmatrix},
    \end{split}
\end{equation}
that is an anti-symmetric ($\eta^{T} \neq \eta$), non-Hermitian matrix ($\eta^{\dagger} \neq \eta$), of null trace, null determinant, with real eigenvalues and that have the properties
\begin{equation}\label{e:ARDE13}
    \begin{split}
        \eta^{2} = (\eta^{\dagger})^{2}=0 \;\; \mbox{and} \;\; \{ \eta, \eta^{\dagger} \} = 2 \mathcal{I}_{4}.
    \end{split}
\end{equation}
The particular one-dimensional case of Eq. (\ref{e:ARDE9}) can then be rewritten as
\begin{equation}
\label{e:ARDE14}
    \pm  \left\{ i \hbar c \alpha_{3} \partial_{3} \right\} \varphi (x^{3}) + \left\{ E_{1} \eta + E_{2} \eta^{\dagger} \right\} \varphi (x^{3}) = 0,
\end{equation}
where $E_{1} = (E - mc^{2})/\sqrt{2}$ and $E_{2} = (E + mc^{2})/\sqrt{2}$. 

On the other hand, Eq. (\ref{e:ARDE9}),  with $j=l=3$, written in terms of the alternative representation, allows obtaining the continuity equation. To do this, we multiply Eq. (\ref{e:ARDE9}) by $\psi^{\dagger}(x^{l})\alpha_{3} \beta$ by the left, obtaining
\begin{align}
    & i \hbar \psi^{\dagger} (x^{l}) \alpha_{3} \eta \eta^{\dagger} \partial_{0} \psi (x^{l}) + i \hbar \psi^{\dagger} (x^{l})\mathcal{I}_{4} \partial_{0} \psi (x^{l}) \label{e:ARDE15} \\
    &-i \hbar c \psi^{\dagger} (x^{l}) \alpha_{3} \beta \alpha_{3} \partial_{3} \psi (x^{l}) - mc^{2} \psi^{\dagger} (x^{l}) \mathcal{I}_{4} \psi (x^{l}) = 0. \nonumber 
\end{align}
Then, we calculate the complex conjugate of Eq. (\ref{e:ARDE9}), and we multiply it by $\beta \alpha_{3} \psi(x^{l})$ on the right hand side to get
\begin{align}
   &-i \hbar \psi^{\dagger} (x^{l}) \eta \eta^{\dagger} \alpha_{3}  \psi (x^{l}) - i \hbar [\partial_{0} \psi^{\dagger} (x^{l})] \mathcal{I}_{4} \partial_{0} \psi (x^{l}) \label{e:ARDE16} \\
   &+ i \hbar c [\partial_{3} \psi^{\dagger} (x^{l})] \alpha_{3} \beta \alpha_{3} \psi (x^{l}) - mc^{2} \psi^{\dagger} (x^{l}) \mathcal{I}_{4} \psi (x^{l}) = 0. \nonumber
\end{align}
By subtracting this last equation from Eq.  (\ref{e:ARDE15}), the continuity equation is finally obtained:
\begin{equation}\label{e:ARDE17}
    \partial_{t}\rho(x^{l}) + \partial_{3} J (x^{l})  = 0.
\end{equation}
Above $\rho(x^{l})=\psi^{\dagger}(x^{l}) \alpha_{3} \psi(x^{l})$ is the probability density while  $J (x^{l})=\psi^{\dagger} (x^{l}) c \alpha_{3} \beta \alpha_{3} \psi (x^{l})$ is the probability current density.
%%%%%\--------------------------------------------------------\%%%%%

%%%%%\--------------------------------------------------------\%%%%%
%%%%%\----------\Sec.III:TheoremWSEAR\------------------------\%%%%%
%%%%%\--------------------------------------------------------\%%%%%
\section{General Theorem of Wave Function Sensitivity to Energy Variations in the Alternative Representation}
\label{sec:sensitivity}

In this section, various algebraic procedures are applied to Eq. (\ref{e:ARDE14}) in order to derive the expression for the general sensitivity theorem of the wave function to energy variations, in a  representation alternative to the one presented by Winful et al.  \cite{Winful200400}. First, we take the derivative of Eq. (\ref{e:ARDE14}), with negative sign, with respect to energy, and we multiply it on the left by $\varphi^{\dagger}(x^{3})$, obtaining the following expression
\begin{align}\label{e:GTWFSEV18}
    &  - \varphi^{\dagger}(x^{3}) \left\{ i \hbar c \alpha_{3} \partial_{3} \right\} \partial_{E} \varphi (x^{3}) = - \varphi^{\dagger}(x^{3}) \left\{ \frac{\eta + \eta^{\dagger}}{\sqrt{2}} \right\} \varphi (x^{3}) \nonumber \\
    &- \varphi^{\dagger}(x^{3}) \{ E_{1} \eta + E_{2} \eta^{\dagger} \} \partial_{E} \varphi (x^{3}).
\end{align}
Secondly, we take the conjugate transpose of Eq. (\ref{e:ARDE14}), with a positive sign, and multiply this result by $\partial_{E} \varphi(x^{3})$. Thus, we obtain
\begin{align}
    & - \left\{ i \hbar c \partial_{3} \varphi^{\dagger}(x^{3}) \alpha_{3} \right\} \partial_{E} \varphi (x^{3})= \nonumber\\
    & - \varphi^{\dagger}(x^{3}) \{ E_{1} \eta + E_{2} \eta^{\dagger} \}^{\dagger} \partial_{E} \varphi (x^{3}). \label{e:GTWFSEV19}
\end{align}
By subtracting equations (\ref{e:GTWFSEV18}) and (\ref{e:GTWFSEV19}), we obtain
\begin{align}
    & -i \hbar c \partial_{3} \left\{ \varphi^{\dagger}(x^{3}) \alpha_{3} \partial_{E} \varphi (x^{3}) \right\} = - \varphi^{\dagger}(x^{3}) \left\{ \frac{\eta + \eta^{\dagger}}{\sqrt{2}} \right\} \varphi (x^{3}) \nonumber \\
    &  - \varphi^{\dagger}(x^{3}) \biggl[ \left\{ E_{1} \eta + E_{2} \eta^{\dagger} \right\}  + \left\{ E_{1} \eta + E_{2} \eta^{\dagger} \right\}^{\dagger} \biggr] \partial_{E} \varphi (x^{3}), \label{e:GTWFSEV20}
\end{align}
    which allows to analyze two cases of interest in the present work for the expressions of energy. In the first case, the rest energy value is considered predominant, ($E_{j} \approx (-1)^{j} mc^{2}$ with $j=1,2$), which leads to
    \begin{equation}\label{e:GTWFSEV21}
        \begin{split}
            E_{1} \eta + E_{2} \eta^{\dagger}
            &=
            \sqrt{2} mc^{2}
            \begin{pmatrix}
                \mathcal{O}_{2} &   -\sigma_{3}    \\
                \sigma_{3}     & \mathcal{O}_{2}
            \end{pmatrix}.
        \end{split}
    \end{equation}
Therefore, Eq. (\ref{e:GTWFSEV21}) implies the equality
    \begin{equation}\label{e:GTWFSEV22}
        \left\{ E_{1} \eta + E_{2} \eta^{\dagger} \right\}^{\dagger} =- \left\{ E_{1} \eta + E_{2} \eta^{\dagger} \right\},
    \end{equation}
    and allows us to rewrite Eq. (\ref{e:GTWFSEV20}) as
    \begin{equation}\label{e:GTWFSEV23}
        -i \hbar c \partial_{3} \left\{ \varphi^{\dagger}(x^{3}) \alpha_{3} \partial_{E} \varphi (x^{3}) \right\} = - \varphi^{\dagger}(x^{3}) \left\{ \frac{\eta + \eta^{\dagger}}{\sqrt{2}} \right\} \varphi (x^{3}).
    \end{equation}
    Consequently, if the expression (\ref{e:GTWFSEV23}) is integrated in the interval from $0$ to $a$, we obtain
    \begin{equation}\label{e:GTWFSEV24}
        -i \hbar c \left\{ \varphi^{\dagger}(x^{3}) \alpha_{3} \partial_{E} \varphi (x^{3}) \right\} \Biggl|_{0}^{a} = - \int_{0}^{a} dx^{3} \varphi^{\dagger}(x^{3}) \beta \varphi (x^{3}).
    \end{equation}
   
    The second case analyzed in this work is obtained by considering the particles energy being large with respect to its rest energy ($E \approx E_{j}$ with $j=1,2$), which allows us to write
    \begin{equation}\label{e:GTWFSEV25}
        E_{1} \eta + E_{2} \eta^{\dagger} \approx E \left\{ \eta + \eta^{\dagger} \right\},
    \end{equation}
    and consequently
    \begin{equation}\label{e:GTWFSEV26}
        \{ E [\eta + \eta^{\dagger}] \}^{\dagger} = E \{ \eta + \eta^{\dagger} \}.
    \end{equation}
We then repeat the procedure applied to obtain Eq. (\ref{e:GTWFSEV20}), but in this case we consider only the expression (\ref{e:ARDE14}) with a positive sign and Eq. (\ref{e:GTWFSEV26}). It can be obtained again the equation corresponding to Eq. (\ref{e:GTWFSEV23}), and consequently (\ref{e:GTWFSEV24}), which is the general sensitivity theorem of the wave function to energy variations, but in the alternative representation. Clearly, the expression (\ref{e:GTWFSEV24}) obtained by considering large energy regimes ($E \gg mc^{2} \rightarrow E \approx E_{j}$) is valid only for massive particles (as in the first case) and would also imply the appearance of the Klein tunneling phenomenon \cite{Winful200400}.
%%%%%\--------------------------------------------------------\%%%%%

%%%%%\--------------------------------------------------------\%%%%%
%%%%%\----------\Sec.IV:SolvingDEAR\--------------------------\%%%%%
%%%%%\--------------------------------------------------------\%%%%%
\section{Solving Dirac's Equation In the Alternative Representation}
\label{sec:solution}

In this section, we obtain exact relationships among the tunneling times, similar to the ones obtained by the authors in Ref. \cite{Winful200400}. We then study  the the manifestation of Hartman's effect in this case. 
%phenomenon of saturation of tunneling times is presented and prove what would be 
Besides, we give the correct expressions that should be obtained for the tunneling times when considering the non-relativistic limit, correcting thus the typos that appear in Ref. \cite{Winful200400}. 

\begin{figure}[t!]
        \centering
        \includegraphics[scale=0.35]{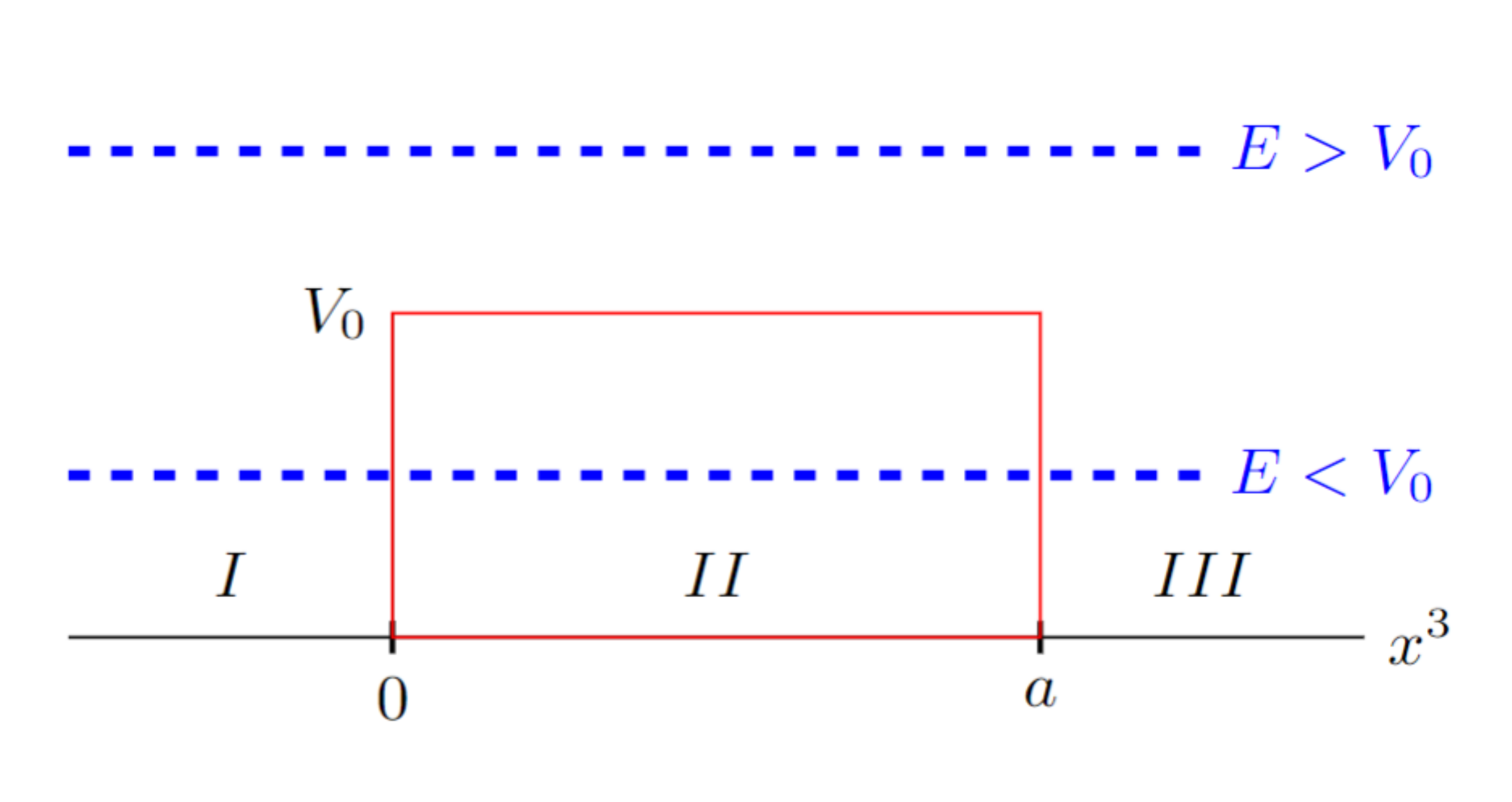}
        \caption{Schematic representation of a potential barrier with width $a$ and height $V_{0}$ for a particle with energy $E$. In the figure, we identify the regions $I$, $II$ and $III$.}
        \label{fig:Quantum tunneling - 00}
    \end{figure}

We start by writing  Eq. (\ref{e:ARDE14}) as
\begin{equation}
\label{e:SDEAR27}
\begin{split}
\begin{pmatrix}
    \mathbf{0}_{2} &   i \hbar c \sigma^{3} \partial_{3}    \\
    i \hbar c \sigma^{3} \partial_{3}    &   \mathbf{0}_{2}
\end{pmatrix}
\begin{pmatrix}
    \varphi_{l} \\
    \varphi_{s}
\end{pmatrix} =
    \begin{pmatrix}
        E \mathbf{1}_{2} &   mc^{2} \sigma^{3}    \\
        -mc^{2} \sigma^{3} & - E \mathbf{1}_{2}
    \end{pmatrix}
    \begin{pmatrix}
        \varphi_{l} \\
        \varphi_{s}
    \end{pmatrix},
\end{split}
\end{equation}
with $\varphi_{l} \equiv \varphi_{l}(x^{3})$ and $\varphi_{s} \equiv \varphi_{s}(x^{3})$. By its turn, this matrix equation can be recast as the following scalar equations:
\begin{align}
        & \varphi_{s}^{j}(x^{3}) = (-1)^{1-j} \frac{1}{E}\left \{ i \hbar c \partial_{3} \varphi_{l}^{j}(x^{3}) + mc^{2} \varphi_{l}^{j}(x^{3}) \right\}, \label{e:SDEAR28} \\
    %\end{equation}
    %and
    %\begin{equation}
        & \frac{1}{\hbar^{2}c^{2}} \left\{ m^{2}c^{4} - E^{2} \right\} \varphi_{l}^{j}(x^{3}) + \partial_{33} \varphi_{l}^{j}(x^{3}) = 0, \label{e:SDEAR29}
    \end{align}
with $j=1,2$ indicating the $j$-th component of $\varphi_l$ or $\varphi_s$.
%, which are obtained by decoupling the system offered by (\ref{e:ARDE14}). In this sense,
We then consider the problem of particles incident from the left on a constant potential barrier, as illustrated in Fig. \ref{fig:Quantum tunneling - 00}, with height $V_{0}$ and width $a$ defined by
\begin{equation}\label{e:SDEAR30}
        V(x^{3})=V_{0} \Theta(x^{3}) \Theta(a-x^{3}),
    \end{equation}
where $\Theta$ is the Heaviside function defined as
    \begin{equation}\label{e:SDEAR31}
        \Theta(x^{3}) = 
        \left\{
            \begin{aligned}
                0 & \;\; \mbox{for} \;\; x^{3} < 0 \\
                1 & \;\; \mbox{for} \;\; x^{3} \leq 0
            \end{aligned}.
        \right.
    \end{equation}

When solving Eq. (\ref{e:SDEAR29}), considering $j=1$ (stationary solution of the particle scattering problem), the component $l$ can be writen as:
\begin{equation}\label{e:SDEAR32}
    \varphi_{l}^{1}(x^{3}) = Ae^{ikx^{3}} + Be^{-ikx^{3}},
\end{equation}
where $A$ and $B$ are coefficients to be determined. Then, the $s$ component is obtained by substituting Eq. (\ref{e:SDEAR32}) into Eq. (\ref{e:SDEAR28}). Thus we get
\begin{equation}\label{e:SDEAR33}
    \varphi_{s}^{1}(x^{3}) = \Gamma_{1} Ae^{ikx^{3}} + \Gamma_{2} Be^{-ikx^{3}},
\end{equation}
where $\Gamma_{1} = mc^{2}/E - \hbar c k/E$ and $\Gamma_{2} = mc^{2}/E + \hbar c k/E$.

The expressions (\ref{e:SDEAR32}) and (\ref{e:SDEAR33}) allow us to write the general solution for each of the regions of interest as
    \begin{align}
            \varphi_{I}(x^{3})
            &=
            A
            \begin{pmatrix}
                1 \\
                0 \\
                -\Gamma \\
                0
            \end{pmatrix}
            e^{ikx^{3}}
            + B
            \begin{pmatrix}
                1 \\
                0 \\
                \Gamma \\
                0
            \end{pmatrix}
            e^{-ikx^{3}}, \label{e:SDEAR34} \\
         \varphi_{II}(x^{3})
            &=
            C
            \begin{pmatrix}
                1 \\
                0 \\
                i\Gamma' \\
                0
            \end{pmatrix}
            e^{qx^{3}}
            + D
            \begin{pmatrix}
                1 \\
                0 \\
                -i\Gamma' \\
                0
            \end{pmatrix}
            e^{-qx^{3}}, \label{e:SDEAR35} \\
            \varphi_{III}(x^{3})
                &=
                F
                \begin{pmatrix}
                    1 \\
                    0 \\
                    -\Gamma \\
                    0
                \end{pmatrix}
                e^{ik(x^{3}-a)}, \label{e:SDEAR36} 
        \end{align}
    where the coefficients $\Gamma$ and $\Gamma^{\prime}$ are obtained from Eq. (\ref{e:SDEAR33}) when considering the lowest energy regimes ($\Gamma_{1} \approx - \hbar c k/E = - \Gamma$, $\Gamma_{2} \approx + \hbar c k/E = \Gamma$ with $\hbar c k = \sqrt{m^{2}c^{4} - E^{2}}$ and $\Gamma' = \hbar c q/E'$ with $E'^{2} = 2m^{ 2}c^{4} - (E - V_{0})^{2}$ and $q^{2} = k^{2}(E') < 0$), in which the Klein tunneling phenomenon does not occur ($mc^{2} \gg E$ and $mc^{2} \gg V_ {0}$).
    
Applying the continuity conditions
\begin{align}
& \varphi_{r}(x^{3})\Big|_{x^{3}} = \varphi_{rI}(x^{3})\Big|_{x^{3}}, \label{e:SDEAR37} \\
& d_{x^{3}}\varphi_{r} (x^{3}) \Big|_{x^{3}} = d_{x^{3}}\varphi_{rI} (x^{3})\Big|_{x^{3}},
\end{align}
with $r=\{I,II\}$ and $x^3=\{0,a\}$, we can obtain the expressions for each of the coefficients:
\begin{align}
    B &= -\frac{i}{2}\frac{(1 + \Xi^{2})}{\Xi} \sinh{\{qa\}} F,
    \label{e:SDEAR38} \\
    C &= \frac{1}{2}(1 + i\Xi) e^{-qa} F,
    \label{e:SDEAR39} \\
    D &= \frac{1}{2}(1 - i\Xi) e^{qa} F,
    \label{e:SDEAR40} \\
    F &= \biggl\{ \cosh{\{qa\}} - \frac{i}{2} \biggl[ \Xi - \frac{1}{\Xi} \biggr] \sinh{\{qa\}} \biggr\}^{-1} A,
    \label{e:SDEAR41}
\end{align}
    with $\Xi=kE'/qE$.    
    For our purposes here, we can assume, without loss of generality, that $A=1$. Using $|T|^{2} + |R|^{2} = 1$ we can rewrite the expressions $B=R=|R|\exp{i\phi_{R}}$ and $F=T=|T|\exp{i\phi_{T}}$. The above considerations allow us to write Eqs. (\ref{e:SDEAR34}), (\ref{e:SDEAR35}), and (\ref{e:SDEAR36}) in terms of the reflection, $R$, and transmission, $T$,  coefficients, which when substituted into Eq. (\ref{e:GTWFSEV24}) gives us
    \begin{align}
        & - 2 \hbar c \Gamma \biggl\{ |T|^{2} \partial_{E}\phi_{T} +  |R|^{2} \partial_{E} \phi_{R} + \frac{1}{\Gamma} Im(R) \partial_{E} \Gamma \biggr\}= \nonumber \\
        &- \frac{a |T|^{2}}{2} \biggl\{ (1 + \Xi^{2})(1 - \Gamma'^{2}) \frac{\sinh2qa}{2qa} + (1 - \Xi^{2})(1 + \Gamma'^{2}) \biggr\}. \label{e:SDEAR42}
    \end{align}
%In the expression (\ref{e:SDEAR44}) 
Definition of the group time for particles transmitted, reflected, self-interaction time and the time group are given by the expressions
    \begin{align}
        \hat{\tau}_{gT} &= \hbar \partial_{E}\phi_{T}, \\
        \hat{\tau}_{gR} &= \hbar \partial_{E}\phi_{R}, \\
        \hat{\tau}_{i}  &= - \hbar Im(R) (\partial_{E} \Gamma)/ \Gamma, \\
        \hat{\tau}_{g}  &= |T|^{2} \hat{\tau}_{gT} + |R|^{2} \hat{\tau }_{gR}.
    \end{align}
From Eq. (\ref{e:SDEAR42}) we can write
\begin{equation}\label{e:SDEAR44}
        %\begin{split}
            |T|^{2} \hat{\tau}_{gT} +  |R|^{2} \hat{\tau}_{gR} - \hat{\tau}_{i}
            =\hat{\tau}_{g} - \hat{\tau}_{i}=\hat{\tau}_{d}.
           % &= \hat{\tau}_{d}.
        %\end{split}
    \end{equation}
Substituting $\Gamma$, $\Gamma'$ and $Im(R)$ into Eq. (\ref{e:SDEAR44}), we obtain  the mathematical expressions for the dwell, interaction and group times in the alternative representation:
\begin{align}
    \hat{\tau}_{d}
            =&
            \frac{a |T|^{2}}{4 \hbar c^{2} q \Xi E'} \Biggl\{ m^{2}c^{4}(1 + \Xi^{2}) \frac{\sinh{\{2qa\}}}{2qa} \nonumber \\
            &+ [3m^{2}c^{4}-2(V_{0}-E)^{2}](1 - \Xi^{2}) \Biggr\},
            \label{e:SDEAR45} \\
    \hat{\tau}_{i}
            =& \frac{m^{2}c^{4}|T|^{2}}{4 \hbar c^{2} k^{2} \Xi E} (1 + \Xi^{2}) \sinh{\{2qa\}},
            \label{e:SDEAR46} \\
    \hat{\tau}_{g}
            =&
            \frac{a |T|^{2}}{4 \hbar c^{2} q \Xi E'} \Biggl\{ m^{2}c^{4}(1 + \Xi^{2}) \frac{\sinh{\{2qa\}}}{2qa} \nonumber \\
            &+ [3m^{2}c^{4}-2(V_{0}-E)^{2}](1 - \Xi^{2}) \Biggr\} \nonumber \\
            &+ \frac{m^{2}c^{4}|T|^{2}}{4 \hbar c^{2} k^{2} \Xi E} (1 + \Xi^{2}) \sinh{\{2qa\}}.
            \label{e:SDEAR47}
\end{align}

\begin{figure}[]
        \centering
        \includegraphics[scale=0.58]{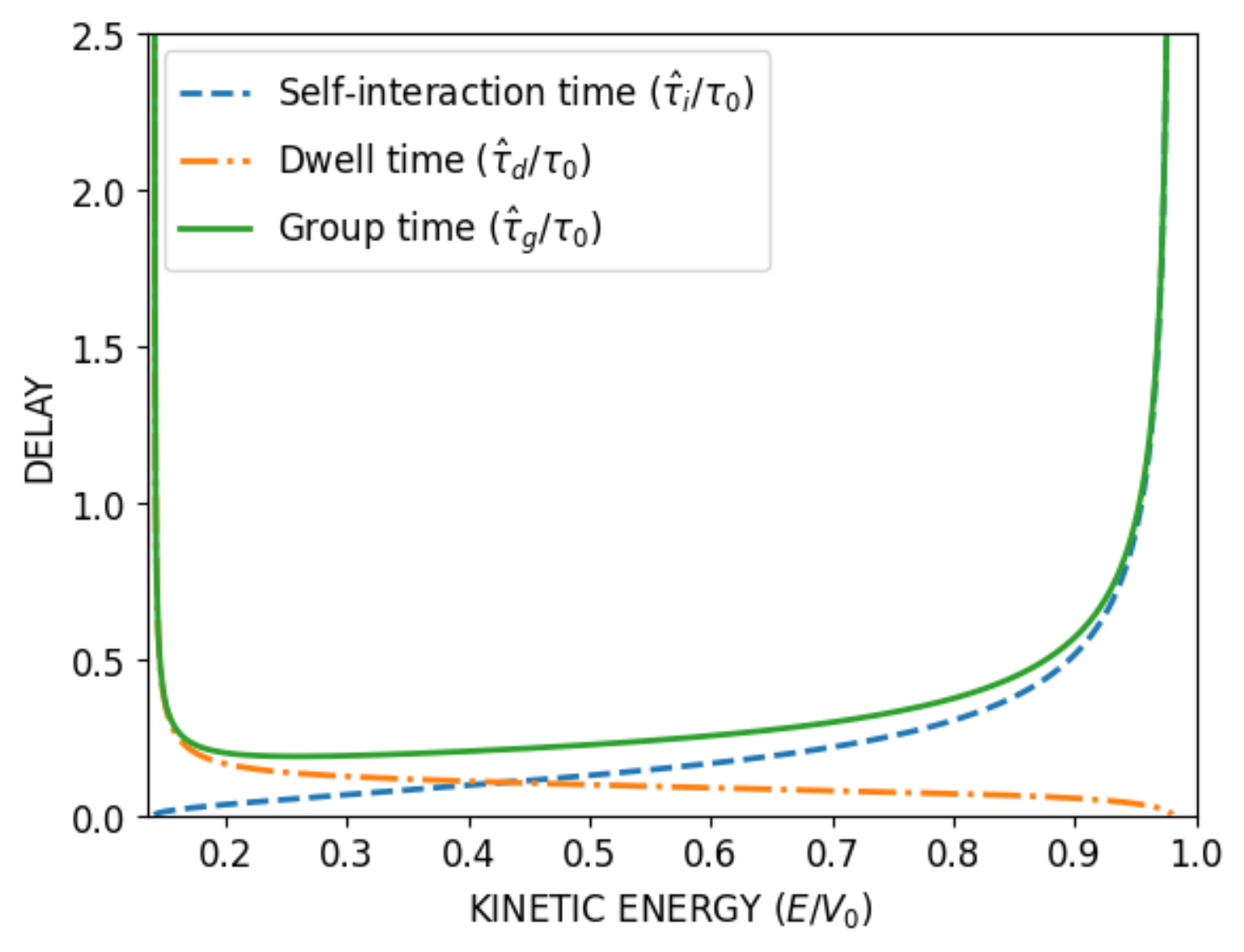}
        \caption{Group time, $\hat{\tau}_{g}/\tau_{0}$, dwell time, $\hat{\tau}_{d}/\tau_{0}$, and self-interaction time, $\hat{\tau}_{i}/\tau_{0}$, as a function of the ratio of energies $E/V_{0}$. The expressions for the times are normalized with respect to the transmission time at  the vacuum light speed, $\tau_{0} = a/c$. Besides, for these plots, we set $V_{0}a/\hbar c = 2 \pi$ and $m c^{2}/V_{0} = 0.98$.}
        \label{fig:Solving the Dirac Equation In This Alternative Representation - 00}
    \end{figure}

When considering very wide potential barriers ($a \rightarrow \infty$), these equations are reduced to the expressions
\begin{align}
    \hat{\tau}_{d} &= \frac{m^{2}c^{4}}{\hbar c^{2} q^{2} E'} \frac{ \Xi}{ \left(1 + \Xi^{2} \right)},
    \label{e:SDEAR48} \\
    \hat{\tau}_{i} &= \frac{2m^{2}c^{4}}{ \hbar c^{2} k^{2} E} \frac{ \Xi}{ \left(1 + \Xi^{2} \right)},
    \label{e:SDEAR49} \\
    \hat{\tau}_{g} &= \frac{m^{2}c^{4}}{\hbar c^{2}} \frac{ \Xi}{ \left(1 + \Xi^{2} \right)} \biggl( \frac{1}{q^{2} E'} + \frac{2}{k^{2} E} \biggr).
    \label{e:SDEAR50}
\end{align} 
The graphical representation of these expressions, shown in Fig. \ref{fig:Solving the Dirac Equation In This Alternative Representation - 00}, allows us to see how the group time curve, $\hat{\tau}_{g}$, is composed through the sum of the curve of the dwell time of the particles inside the barrier, $\hat{\tau}_{d}$, with the curve of the interaction time between the incident and reflected wave pulses, $ \hat{\tau}_{i}$.
%, as deduced by the authors \cite{ Winful200300,Winful200400}. 
In addition, the group time saturation
%, $\hat{\tau}_{g}$, 
is seen when $E/V_{0}$ goes toward $1$. However, it is important to point out that in the alternative representation,  Eqs. (\ref{e:ARDE10}) and (\ref{e:ARDE11}), the time scales are altered when compared the the ones obtained in Ref. \cite{Winful200400}, since here the interaction time, $\hat{\tau}_{i}$, and dwell time, $\hat{\tau}_{d}$, are modulated by $E^{-1}$ and $(2E^{\prime})^{ -1}$, respectively. This causes a distortion in the group time curve, that is, a more pronounced growth of $\hat{\tau}_{g}$ when $E/V_{0} \rightarrow 1.0$, which can be attributed to the loss of symmetry associated with the matrix $\eta$. On the other hand, it is evident that, as a result of the chosen representation, there is an exchange in the behaviors of the dwell and self-interaction times in relation to what was reported in Ref. \cite{Winful200400}.
    
\begin{figure}[]
        \centering
        \includegraphics[scale=0.58]{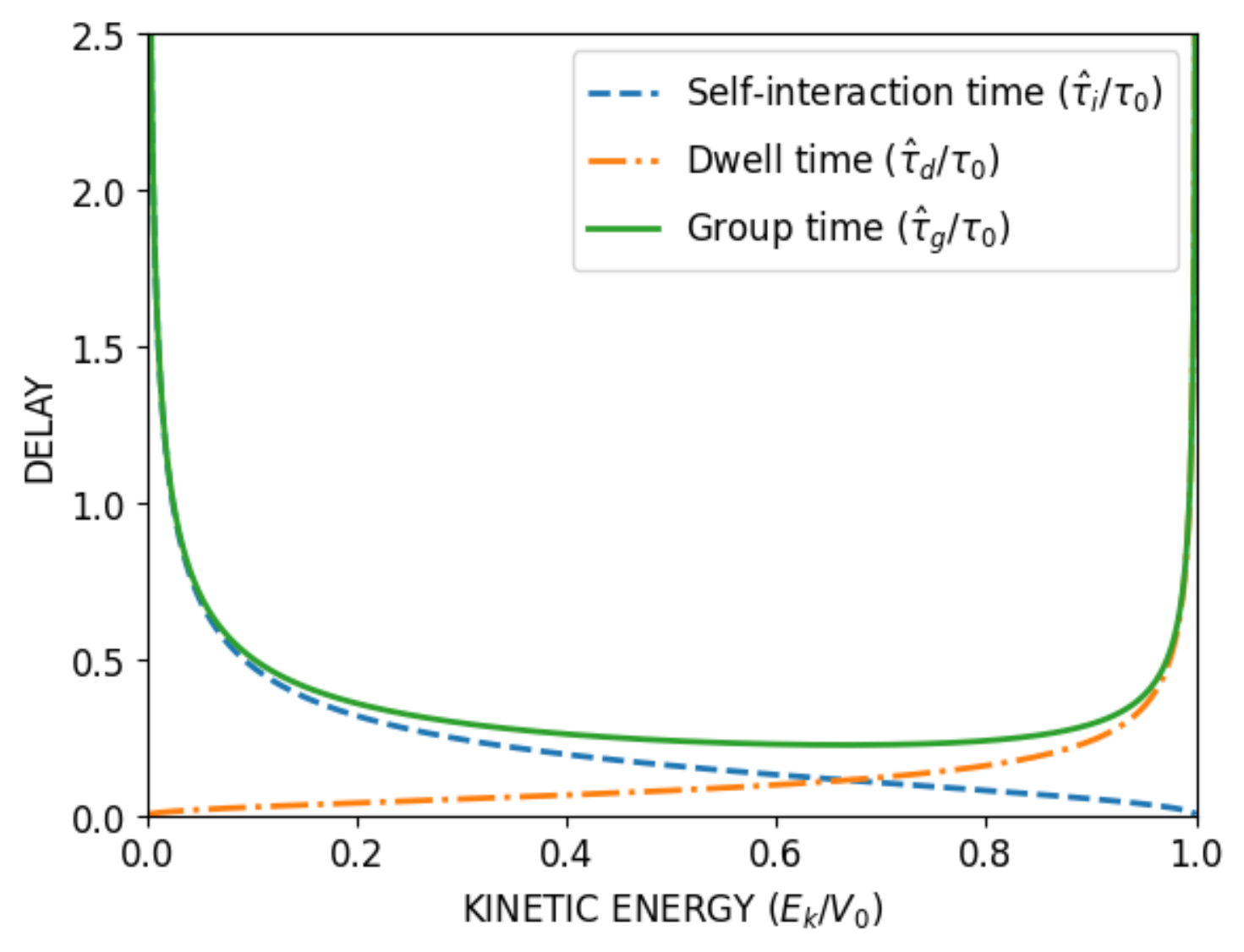}
        \caption{Group time, $\hat{\tau}_{g}/\tau_{0}$, dwell time, $\hat{\tau}_{d}/\tau_{0}$, and self-interaction time, $\hat{\tau}_{i}/\tau_{0}$, as a function of the energies ratio  $E_{k}/V_{0}$ in the non-relativistic limit. The expressions for the times are normalized with respect to the transmission time for the vacuum light speed, $\tau_{0} = a/c$. Besides, for these plots, we set $V_{0}a/\hbar c = 2 \pi$.}
        \label{fig:Solving the Dirac Equation In This Alternative Representation - 01}
    \end{figure}

The non-relativistic limit of the expressions (\ref{e:SDEAR45}), (\ref{e:SDEAR46}) and (\ref{e:SDEAR47}), is obtained by substituting $E = \sqrt{2} E_{1} =E_{k}-mc^{2}$ in $\hbar c k = \sqrt{m^{2}c^{4}- E^{2}}$, $E = \sqrt{2} E_{2} =E_{k}+mc^{2}$ in $\hbar c q = \sqrt{m^{ 2}c^{4}- (V_{0}-E)^{2}}$ and $E'=\sqrt{2m^{2}c^{4}-(V_{0}-E)^ {2}}$. This allows us to write $k' = \sqrt{2mE_{k}}/\hbar$, $q' = \sqrt{2m(V_{0}-E_{k})}/\hbar$, $E \approx mc^{2}$ and $E' = mc^{2} \Rightarrow \Xi = k'/q'$. Consequently
  \begin{align}
        \hat{\tau}_{i}
                =&
                \frac{m|T'|^{2}a}{2 \hbar k'} \left(1 + \frac{q'^{2}}{k'^{2}} \right) \frac{\sinh{\{2q'a\}}}{2q'a},
        \label{e:SDEAR51} \\
        \hat{\tau}_{d}
                =&
                    \frac{m |T'|^{2} a}{4 \hbar k'} \Biggl\{\left(1 + \frac{k'^{2}}{q'^{2}}\right) \frac{\sinh{\{2q'a\}}}{2q'a} + \left(1 - \frac{k'^{2}}{q'^{2}}\right) \Biggr\},
        \label{e:SDEAR52} \\
        \hat{\tau}_{g}
                =&
                \frac{m |T^{\prime}|^{2}a}{2 \hbar k^{\prime}} \Biggl\{ \Biggl[\frac{1}{2} \left( 1 + \frac{k^{\prime 2}}{q^{\prime 2}} \right) \nonumber \\
                &+ \left( 1 + \frac{q^{\prime 2}}{k^{\prime 2}} \right) \Biggr]\frac{\sinh{\{2q^{\prime}a\}}}{2q^{\prime}a} + \left(1 - \frac{k^{\prime 2}}{q^{\prime 2}}\right) \Biggr\},
        \label{e:SDEAR53}
    \end{align}
where $T^{\prime}=\{ \cosh{\{q^{\prime}a\}} - (i/2)[ k^{\prime}/q^{\prime} - q^{\prime}/k^{\prime} ] \sinh{\{q^{\prime}a\}} \}^{-1}$. Thus, when considering very wide barriers ($a \rightarrow \infty$), these equations are reduced to
    \begin{align}
        \hat{\tau}_{d} &= \frac{m \Xi}{\hbar q'^{2} \left(1 + \Xi^{2} \right)},
        \label{e:SDEAR54} \\
        \hat{\tau}_{i} &= \frac{2m \Xi}{\hbar k'^{2} \left(1 + \Xi^{2} \right)},
        \label{e:SDEAR55} \\
        \hat{\tau}_{g} &= \frac{m \Xi}{\hbar \left(1 + \Xi^{2} \right)} \biggl( \frac{1}{q'^{2}} + \frac{2}{k'^{2}} \biggr).
        \label{e:SDEAR56}
    \end{align}
The graphical representation of the expressions (\ref{e:SDEAR54}), (\ref{e:SDEAR55}) and (\ref{e:SDEAR56}) is given in Fig. \ref{fig:Solving the Dirac Equation In This Alternative Representation - 01}. We see how the group time curve, in the non-relativistic limit, is again composed by the sum of the time curve of the dwell time and of the interaction time curve between, as in the model deduced in Ref. \cite{ Winful200300,Winful200400}. 
%Where the green curve 
Besides, Fig. \ref{fig:Solving the Dirac Equation In This Alternative Representation - 01} shows again the saturation effect of the group time, when $E_{k}/V_{0}$ goes toward $1$. Furthermore, it is important to point out the shift to the right we observe in the minimum of the curve of the group time, when $E_{k}/V_{0} \rightarrow 1.0$. This effect is caused by the halving (in orders of magnitude) of the dwell time $\hat{\tau}_{d}$ with respect to the dwell time $\tau_{d}$ obtained in Ref. \cite{Winful200400}. This effect can be attributed, in the same way as the relativistic case, to the loss of symmetry associated with the matrix $\eta$, that was used in the alternative representation for Dirac's equation we use here. On the other hand, in the non-relativistic limit, the dwell time $\hat{\tau}_{d}$ and interaction time $\hat{ \tau}_{i}$ keep the same behavior they had in the representation chosen in Ref. \cite{Winful200400}.

In contrast to the expressions (\ref{e:SDEAR51}), (\ref{e:SDEAR52}) and (\ref{e:SDEAR53}), in Ref. \cite{Winful200400}, the authors present a set of mathematical expressions for the non-relativistic case, from which it is clearly not possible to recover the characteristic saturation curve of tunneling times (the Hartman effect) when considering a very wide potential barrier. However, such non-relativistic expressions would be correct if the quantity $m |T^{\prime}|^{2} a/2 \hbar k^{\prime}$\footnote{To clarify the difference in nomenclature used in our work with respect to that used in Ref. \cite{Winful200400}, we can define $k^{\prime} = k$, $f^{-1} = | T^{\prime} |$, $q^{\prime} = \kappa$ and $L = a$.} multiplied all the terms that define the phase time $\tau_{g}$ and the dwell time $\tau_{d}$, and if, in addition, the first term in brackets in $\tau_{g}$ were squared (see Ref. \cite{Winful200300,Winful200600}).

\section{Influence of Spin on Tunneling Times}
\label{sec:spin}

By considering the solution of Dirac's equation (Eqs. (\ref{e:SDEAR34}), (\ref{e:SDEAR35}) and (\ref{e:SDEAR36})) associated with the lower energy regime where Klein tunneling is not present, in Sec. \ref{sec:solution} we showed that the alternative representation for Dirac's equation (Eqs. (\ref{e:ARDE10}) and (\ref{e:ARDE11})) leads to exact relationships between tunneling times similar to those reported in Ref. \cite{Winful200400}. However, the most relevant aspect of this alternative representation is obtained by writing Eq. (\ref{e:ARDE14}), with the negative sign, in the momentum basis. That is, by considering $\hat{P} = -i\hbar \partial_{3}$ and $\varphi = \varphi' (p^{l})\exp{-i \mathbf{p} \Lambda \mathbf{x}}=\varphi' (p^{l})\exp{-i[Et - p^{l}x^{l}]}=\varphi (p^{l})\exp{- iEt}$\footnote{In this article, we apply the usual designation $\vec{p} = (p_{x}, p_{y}, p_{z})$ for the linear momentum and $E$ for the energy in the $4$-momentum $\textbf{p}$. Also, we consider the coordinate transformation $\textbf{x}'=\Lambda \textbf{x}=\Lambda (ct,\vec{x})^{T}$, where $\Lambda$ has matrix elements $\Lambda^{0}_{\ 0} = \Lambda^{1}_{\ 1} = \gamma$, $\Lambda^{0}_{\ 1} = \Lambda^{1}_{\ 0} = - \gamma v$, $\Lambda^{0}_{\ 2} = \Lambda^{2}_{\ 0} = \Lambda^{0}_{\ 3} = \Lambda^{3}_{\ 0} = \Lambda^{2}_{\ 1} = \Lambda^{1}_{\ 2} = \Lambda^{3}_{\ 1} = \Lambda^{1}_{\ 3} = \Lambda^{2}_{\ 3} = \Lambda^{3}_{\ 2} = 0$ and belongs to the Lorentz group, $\mathcal{L}$, and leaves the quadratic form $t^{2} - |\vec{x}|^{2}$ invariant. The matrix associated with this quadratic form is $\zeta = diag(1,-1,-1,-1)$, which satisfies the relation $\Lambda^{T} \zeta \Lambda = \zeta$.}, 
we derive the equations
    \begin{align}
        E_{k} \varphi_{l}^{j}(p^{3}) + (-1)^{1-j} cp \varphi_{s}^{j}(p^{3})  = 0, \label{e:ISTT63} \\
    %\end{equation}
    %and
    %\begin{equation}
        (-1)^{1-j} cp \varphi_{l}^{j}(p^{3}) - E_{k} \varphi_{s}^{j}(p^{3}) = 0, \label{e:ISTT64}
    \end{align}
    with $j = 1,2$. Considering super-relativistic energy regime ($E_{k} \gg mc^{2} \rightarrow E \approx E_{k}= \hbar^{2} p^{2}/2m$ and $cp \pm mc^{2} \approx cp$), in which the Klein's tunneling phenomenon is again not appreciable ($E -mc^{2} < V_{0} < E + mc^{2}$), we solve Eqs. (\ref{e:ISTT63}) and (\ref{e:ISTT64}) to obtain the general solution for each of the regions:
    \begin{align}
        \varphi_{I}(x^{3})
            &= \frac{A}{\Upsilon^{2}}
            \begin{pmatrix}
                1 \\
                0 \\
                \Gamma \\
                0
            \end{pmatrix}
            e^{ipx^{3}}
            +
            \frac{A'}{\Upsilon^{2}}
            \begin{pmatrix}
                1 \\
                0 \\
                -\Gamma \\
                0
            \end{pmatrix}
            e^{-ipx^{3}} \nonumber \\
            &+
            \frac{B'}{\Upsilon^{2}}
            \begin{pmatrix}
                0 \\
                1 \\
                0 \\
                -\Gamma
            \end{pmatrix}
            e^{-ipx^{3}},
            \label{e:ISTT65} \\
            \varphi_{II}(x^{3})
            &=
            \frac{C}{\Upsilon'^{2}}
            \begin{pmatrix}
                1 \\
                0 \\
                i\Gamma' \\
                0
            \end{pmatrix}
            e^{-p'x^{3}}
            +
            \frac{C'}{\Upsilon'^{2}}
            \begin{pmatrix}
                1 \\
                0 \\
                -i\Gamma' \\
                0
            \end{pmatrix}
            e^{p'x^{3}} \nonumber \\
            &   
            +
            \frac{D}{\Upsilon'^{2}}
            \begin{pmatrix}
                0 \\
                1 \\
                0 \\
                i\Gamma'
            \end{pmatrix}
            e^{-px^{3}}
            +
            \frac{D'}{\Upsilon'^{2}}
            \begin{pmatrix}
                0 \\
                1 \\
                0 \\
                -i\Gamma
            \end{pmatrix}
            e^{p'x^{3}},
            \label{e:ISTT66} \\
            \varphi_{III}(x^{3})
            &= \frac{F}{\Upsilon^{2}}
            \begin{pmatrix}
            1 \\
            0 \\
            \Gamma \\
            0
        \end{pmatrix}
            e^{ip(x^{3}-a)}
            +
            \frac{G}{\Upsilon^{2}}
            \begin{pmatrix}
            0 \\
            1 \\
            0 \\
            \Gamma
        \end{pmatrix}
            e^{ip(x^{3}-a)},
        \label{e:ISTT67}
    \end{align}
with $\Gamma = cp/E_{k}$, $\Gamma^{\prime} = cp^{\prime}/E^{\prime}_{k}$, $p^{\prime}=\sqrt{2m(V_{0}-E_{k})}/\hbar$, $E^{\prime}_{k}= V_{0} - E_{k}$, $[\Upsilon^{2}]^{-1} = 1/\sqrt{1+(\pm cp)^{2}/ E_{k}^{2}}$ and $[\Upsilon^{\prime 2}]^{-1} = 1/\sqrt{1+(\pm cp^{\prime})^{2}/E_{k}^{\prime 2}}$. Furthermore, it becomes evident that the expressions (\ref{e:ISTT65}), (\ref{e:ISTT66}) and (\ref{e:ISTT67}) are written as a linear combination of the vectors
    \begin{equation}
    \label{e:ISTT68}
        \varphi_{\pm}^{\uparrow} = \frac{1}{\Upsilon^{2}}
            \begin{pmatrix}
            1 \\
            0 \\
            \pm \Gamma \\
            0
            \end{pmatrix}
            \;\; \mbox{and} \;\;
            \varphi_{\pm}^{\downarrow} = \frac{1}{\Upsilon^{2}}
            \begin{pmatrix}
            0 \\
            1 \\
            0 \\
            \pm \Gamma
        \end{pmatrix}.
    \end{equation}
These vectors satisfy the orthogonality relations
    \begin{align}%\label{}
        %\begin{split}
            & [\varphi_{\pm}^{\uparrow}]^{\dagger} \varphi_{\pm}^{\uparrow} = 1, \label{e:ISTT69}\\
       % \end{split}
    %\end{equation}
   % and
   % \begin{equation}\label{}
       % \begin{split}
           & [\varphi_{\pm}^{\uparrow}]^{\dagger} \varphi_{\pm}^{\downarrow} = [\varphi_{\pm}^{\downarrow}]^{\dagger} \varphi_{\pm}^{\uparrow} = [\varphi_{\mp}^{\downarrow}]^{\dagger} \varphi_{\mp}^{\uparrow} = 0.
           \label{e:ISTT70}
       % \end{split}
    \end{align}
    On the other hand, the coefficients $A$, $A'$, $B'$, $C$, $C'$, $D$, $D'$, $F$, and $G$ in Eqs. (\ref {e:ISTT65}), (\ref{e:ISTT66}) and (\ref{e:ISTT67}) are determined by applying the continuity condition
    \begin{equation}\label{e:ISTT71}
        \begin{split}
            \varphi_{r} (x^{3}) \big|_{x^{3}} 
            &= \varphi_{rI} (x^{3}) \big|_{x^{3}},
        \end{split}
    \end{equation}
    with $r=\{I,II$\} and $x^{3}=\{0,a\}$. Thus, we obtain
    \begin{align}
        A'&= -\frac{i}{2}\frac{(1 + \Xi^{2})}{\Xi} \sinh{\{p'a\}} F,
        \label{e:ISTT72} \\
        C &= \frac{\Upsilon'^{2}}{2\Upsilon^{2}}(1 - i\Xi) e^{p'a} F,
        \label{e:ISTT73} \\
        C'&= \frac{\Upsilon'^{2}}{2\Upsilon^{2}}(1 + i\Xi) e^{-p'a} F,
        \label{e:ISTT74} \\
        D &= \frac{\Upsilon'^{2}}{2\Upsilon^{2}}(1 - i\Xi) e^{p'a} G,
        \label{e:ISTT75} \\
        D'&= \frac{\Upsilon'^{2}}{2\Upsilon^{2}}(1 + i\Xi) e^{-p'a} G,
        \label{e:ISTT76} \\
        F &= \biggl\{ \cosh{\{p'a\}} - \frac{i}{2} \biggl[ \Xi - \frac{1}{\Xi} \biggr] \sinh{\{p'a\}} \biggr\}^{-1},
        \label{e:ISTT77} \\
        G &= \biggl\{ \cosh{\{p'a\}} - i \Xi \sinh{\{p'a\}} \biggr\}^{-1} B'.
        \label{e:ISTT78}
    \end{align}
    Having these equations,
    %Eqs. (\ref{e:ISTT72}), (\ref{e:ISTT73}), (\ref{e:ISTT74}), (\ref{e:ISTT75}), (\ref{e:ISTT76}), (\ref{e:ISTT77}) and (\ref{e:ISTT78}), 
    for our purposes here, we can choose, without loss of generality, $A = 1$. So $A' \propto |R| e^{i\phi_{R}}$, $B' \propto |R'| e^{i\phi_{R'}}$, $F \propto |T| e^{i\phi_{T}}$ and $G \propto |T'| e^{i\phi_{T'}}$. Besides $|T|^{2} + |T'|^{2}+|R|^{2}+|R'|^{2} = 1$. With this, we can rewrite Eqs. (\ref{e:ISTT65}), (\ref{e:ISTT66}), and (\ref{e:ISTT67}) in terms of these parameters, and substitute them into Eq. (\ref{e:GTWFSEV24}). With this, we arrive at
    \begin{align}
        & \frac{2 \hbar c \Gamma}{[\Upsilon^{2}]^{2}} \biggl\{ |T|^{2} \partial_{E_{k}} \phi_{T} + |T'|^{2} \partial_{E_{k}} \phi_{T'} \nonumber \\
        & \hspace{1.5cm} + |R|^{2} \partial_{E_{k}} \phi_{R} + |R'|^{2} \partial_{E_{k}} \phi_{R'} - Im(R) \frac{\partial_{E_{k}} \Gamma}{\Gamma} \biggr\} \nonumber \\
        & = - \frac{|T|^{2}a}{2 [\Upsilon^{2}]^{2}} \biggl\{(1 - \Gamma'^{2})(1 + \Xi^{2})\frac{\sinh{(2p'a)}}{2p'a} \nonumber \\
        & \hspace{1.8cm} + (1 + \Gamma'^{2})(1 - \Xi^{2}) \biggr\} \nonumber \\
        & - \frac{|T'|^{2}a}{2 [\Upsilon^{2}]^{2}} \biggl\{(1 - \Gamma'^{2})(1 + \Xi^{2})\frac{\sinh{(2p'a)}}{2p'a} \nonumber \\
        & \hspace{1.5cm} + (1 + \Gamma'^{2})(1 - \Xi^{2}) \biggr\}
        \label{e:ISTT79}.
        %&= \hat{\tau}_{d}^{\uparrow} + \hat{\tau}_{d}^{\downarrow}.
        %\label{e:Influence of Spin on Tunneling Times - 61}
    \end{align}
From this last equation, we identify the tunneling times 
    \begin{align}
        \hat{\tau}_{g}^{\uparrow}   &= |T|^{2} (\partial_{E_{k}} \phi_{T}) + |R|^{2} (\partial_{E_{k}} \phi_{R}), \\
        \hat{\tau}_{g}^{\downarrow} &= |T'|^{2} (\partial_{E_{k}} \phi_{T'}) + |R'|^{2} (\partial_{E_{k}} \phi_{R'}), \\
        \hat{\tau}_{i}              &= \frac{Im(R)(\partial_{E_{k}} \Gamma)}{\Gamma}, \\
        \hat{\tau}_{g}              &= \hat{\tau}_{g}^{\uparrow} + \hat{\tau}_{g}^{\downarrow}.
    \end{align}
Then, by substituting $\Gamma$, $\Gamma^{\prime}$ and $Im(R)$ into Eq. (\ref{e:ISTT79}), we obtain the following expressions for tunneling times
    \begin{align}
        -\frac{2 \hbar p' \Xi}{ m|T|^{2}a}\hat{\tau}_{d}^{\uparrow}
                &=   [ (E_{k} - V_{0})/2mc^{2} - 1](1 + \Xi^{2}) \frac{\sinh{(2p'a)}}{2p'a} \nonumber \\
                &+ [(E_{k} - V_{0})/2mc^{2} + 1](1 - \Xi^{2}),
        \label{e:ISTT80} \\
       -\frac{2 \hbar p' \Xi}{ m|T'|^{2}a} \hat{\tau}_{d}^{\downarrow}
                &=  [(E_{k} - V_{0})/2mc^{2} - 1](1 + \Xi^{2}) \frac{\sinh{(2p'a)}}{2p'a} \nonumber \\
                &+ [(E_{k} - V_{0})/2mc^{2} + 1](1 - \Xi^{2}),
        \label{e:ISTT81} \\
        \hat{\tau}_{i} &=  \frac{m |T|^{2}}{4 \hbar p^{2} \Xi} (1 + \Xi^{2}) \sinh{(2p'a)},
        \label{e:ISTT82} \\
        \hat{\tau}_{g} &= \hat{\tau}_{d}^{\uparrow} + \hat{\tau}_{d}^{\downarrow} + \hat{\tau}_{i}.
        \label{e:ISTT83}
    \end{align}
Now, if we regard very wide barriers ($a \rightarrow \infty$), the expressions above are reduced to
    \begin{align}
        \hat{\tau}^{\uparrow}_{d} &= -\frac{2 m \left[ (E_{k} - V_{0})/2mc^{2} - 1 \right] \Xi}{\hbar p'^{2} \left(1 + \Xi^{2} \right)},
        \label{e:ISTT84} \\
        \hat{\tau}^{\downarrow}_{d} &= -\frac{2 m \left[ (E_{k} - V_{0})/2mc^{2} - 1 \right] \Xi}{\hbar p'^{2} \left(1 + \Xi^{2} \right)},
        \label{e:ISTT85} \\
        \hat{\tau}_{i} &=  \frac{2 m \Xi}{\hbar p^{2} \left(1 + \Xi^{2} \right)},
        \label{e:ISTT86} \\
        \hat{\tau}_{g} &= -\frac{2m \Xi}{\hbar \left(1 + \Xi^{2} \right)} \biggl\{ \frac{\left[ (E_{k} - V_{0})/mc^{2} - 2 \right]}{p'^{2}} - \frac{1}{p^{2}} \biggr\}.
        \label{e:ISTT87}
    \end{align}

\begin{figure}[]
        \centering
        \includegraphics[scale=0.58]{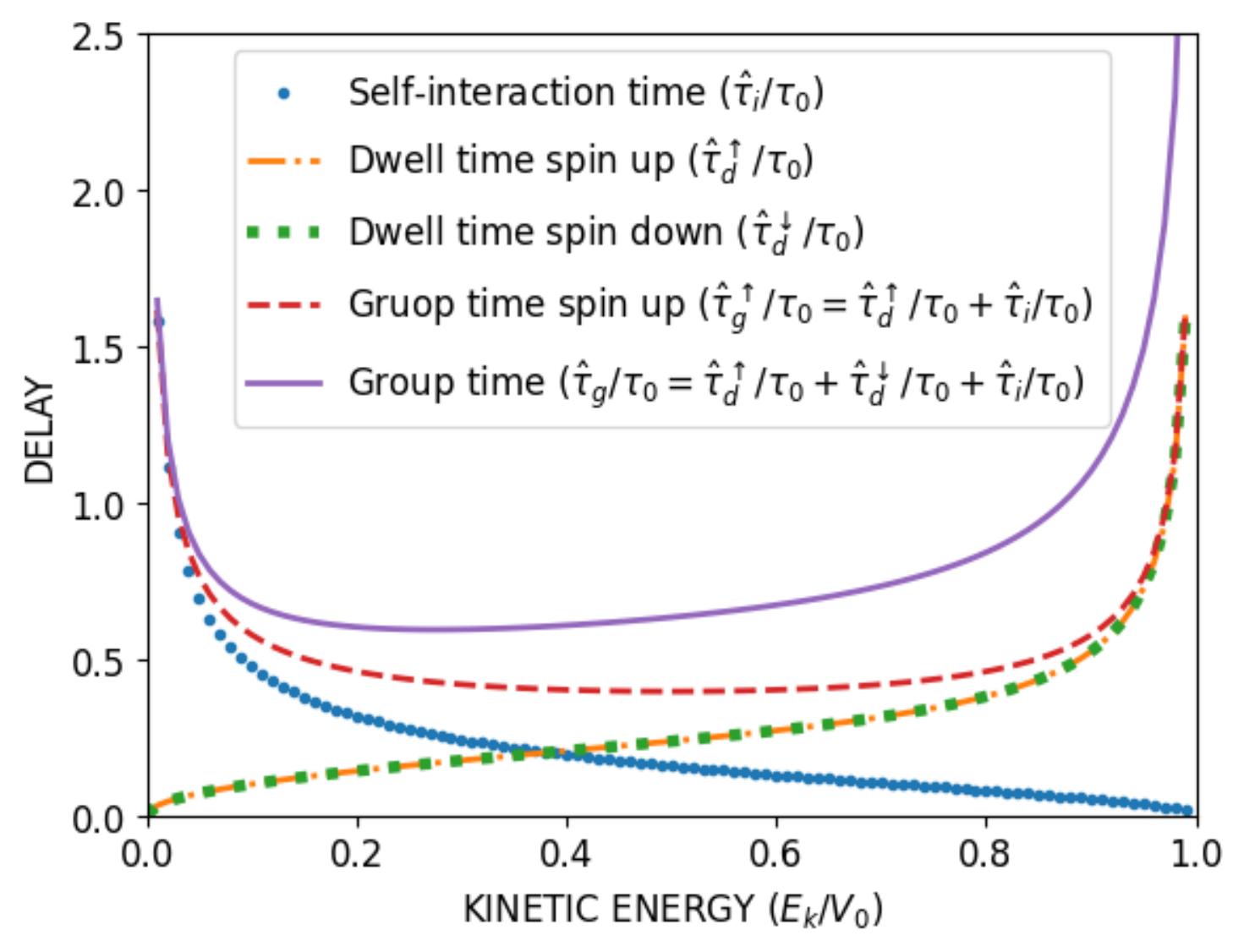}
        \caption{Group time, $\hat{\tau}_{g}/\tau_{0}$, group time considering only the contribution of the spin up particles, $\hat{\tau}_{g}^{\uparrow }/\tau_{0}$, dwell times, $\hat{\tau}_{d}^{\uparrow}/\tau_{0}$ and $\hat{\tau}_{d}^ {\downarrow}/\tau_{0}$, and self-interaction time, $\hat{\tau}_{i}/\tau_{0}$ as a function of the ratio of energies $E_{k}/V_{0}$ in the super-relativistic limit. The expressions for the times are normalized with respect to the transmission time at the vacuum light speed, $\tau_{0} = a/c$. Besides, for these plots, we set $V_{0}a/\hbar c = 2 \pi$ and $V_{0}/mc^{2}=0.98$.}
        \label{fig:Influence of Spin on Tunneling Times - 00}
    \end{figure}

The graphical representation of these expressions is presented in Fig. \ref{fig:Influence of Spin on Tunneling Times - 00}.
%of the expressions (\ref{e:ISTT84}), (\ref{e:ISTT85}), (\ref{e:ISTT86}) and (\ref{e:ISTT87}) 
The curves for tunneling times obtained in the super-relativistic regime are seen to be similar to those obtained in Sec. \ref{sec:solution} for the relativistic and non-relativistic regimes. In other words, by summing the curves for the dwell times of particles with spin up, $\hat{\tau}_{d}^{\uparrow}$, and with spin down, $\hat{\tau}_{d}^{\downarrow}$ and the curve of the interaction time between incident and reflected particles with spin up, $\hat{\tau}_{i}$, a curve for the group time is obtained, $\hat{\tau}_{g}$, which is similar to that obtained in Sec. \ref{sec:solution} for the group time in the relativistic regime. 
On the other hand, it is evident that if the term of the dwell time for particles with spin down, $\hat{\tau}_{d}^{\downarrow}$, is eliminated in that sum, a curve is obtained with the shifted minimum for the group time, $\hat{\tau}_{g}^{\uparrow}$, as obtained in Sec. \ref{sec:solution} for the non-relativistic regime. 
It is interesting that by changing the representation in which we express Eq. (\ref{e:ARDE14}), the results of the saturation curves can be obtained as a consequence of the contribution of the tunneling times of particles with a specific spin, since in the super-relativistic energy regime the reflection of particles with spin down can occur even in the absence of external magnetic fields.
    
%%%%%\--------------------------------------------------------\%%%%%

%%%%%\--------------------------------------------------------\%%%%%
%%%%%\----------\Sec.VI:Conclusion\---------------------------\%%%%%
%%%%%\--------------------------------------------------------\%%%%%
\section{Conclusion}
\label{sec:conclusion}
    In this article, we built a mathematical model to obtain tunneling times through a procedure analogous to the one proposed in Ref. \cite{Winful200400}, but based on an alternative representation of Dirac's equation. This allowed us to find exact relationships between tunneling times, similar to those obtained in Ref. \cite{Winful200400} for relativistic and non-relativistic energy regimes. Moreover, the use of this alternative representation allowed us to analyze the influence of the particle spin on the exact relationships between tunneling times obtained by studying this phenomenon considering a constant potential barrier, super relativistic energy regimes and the absence of external magnetic fields.
    
    From the mathematical expressions obtained in Sec. \ref{sec:spin}, we showed that, in this regime, the group time, $\hat{\tau}_{g}$, is obtained as the sum of the dwell times inside the potential barrier for particles with spin up, $\hat{\tau}_{d}^{\uparrow}$, and with spin down, $\hat{\tau}_{d}^{\downarrow}$, and the self-interaction time, $\hat{\tau}_{i}$, associated with the incident and reflected wave functions for particles with spin up. Indeed, the above summation resulted in a group time curve similar to that obtained in Sec. \ref{sec:solution} for relativistic energy regimes. In addition, it was shown that when considering only the contribution of the particles with spin up, the result obtained is similar to the group time curve presented in Sec. \ref{sec:solution}, with a minimum point shifted to the right and obtained when considering non-relativistic energy regimes.
    
    However, it is interesting that from a purely mathematical procedure, that is, the choice of an alternative representation for Dirac's equation, and the consideration of super relativistic energy regimes, a model is obtained that leads to results involving the influence of particle spin on the exact relationships between tunneling times. Furthermore, it is important to highlight that the results presented in our work depend on the chosen representation, which is based on the one introduced by Ajaib (Refs. \cite{Ajaib201500,Ajaib201600}).
    
    Consequently, it is this representation that leads to the new expressions for the tunneling times, which cannot be obtained by direct application of the Dirac equation written in terms of the standard representation of $\gamma$ matrices. This is so because in such an approach there is no apparent physical or mathematical reason to consider an ansatz as presented in Sec. \ref{sec:spin}. By taking the non-relativistic limit of the Dirac equation in its standard form, the Schr\"{o}dinger equation is obtained in which the influence of spin does not appear in the absence of a magnetic field.
    Therefore, considering an ansatz like the one proposed in Sec. \ref{sec:spin} is interesting only for this type of alternative representation. In addition, for the Dirac equation considered in Sec. \ref{sec:solution}, from its first order is obtained the equation introduced by Ajaib in Refs. \cite{Ajaib201500,Ajaib201600}, where the influence of spin is mathematically appreciable even in the non-relativistic regime and in the absence of an external magnetic field.
    
    On the other hand, the fact that the results obtained depend on the chosen representation seems to lead to a theoretical model that offers the description of different tunneling times. This may seem like a physically inconsistent result, since, in general, any physical result should be independent of the mathematical tool used. However, it should be noted that the formalism of the Dirac equation dictates only that the representations of Dirac's have to be such that they warrant its invariance under a Lorentz transformations and lead to an Hermitian Hamiltonian.
    
    For that reason, there is no impediment for a description of different tunneling times to emerge as a consequence of the representation change. This apparent controversy between what seems to be the description of different types of tunneling times becomes an open problem that can only be resolved by carrying out the experiment itself. In other words, if it is possible to carry out an experimental measurement of the tunneling time (in the absence of an external magnetic field and for systems that can reach super relativistic energy regimes), our theoretical predictions could then be tested.
    %a result consistent with the one that emerges from the representation used in this work could be obtained.
     This is left as an open problem for future research.
    
    Finally, the results reported in this article could, in principle, be applied to nuclear physics phenomena in high energy regimes. However, it is important to highlight that the Hartman effect is present in all our results. Therefore, more future work must be done to properly define the tunneling time of quantum systems.

\begin{acknowledgments}
This work was supported by the National Institute for the Science and Technology of Quantum Information (INCT-IQ), process 465469/2014-0, and by the National Council for Scientific and Technological Development (CNPq), processes 309862/2021-3 and 409673/2022-6.
\end{acknowledgments}
%%%%%\--------------------------------------------------------\%%%%%
%%%%%\----------\Referencias Bibligráficas\-------------------\%%%%%
%%%%%\--------------------------------------------------------\%%%%%

%%%%%\--------------------------------------------------------\%%%%%
\end{document}